\documentclass[final,1p,times]{elsarticle} % do not change
\usepackage{graphicx} % do not change
\usepackage{amssymb} % do not change
\usepackage{amsthm} % do not change
\usepackage{lineno} % do not change
\journal{Nuclear Physics A} % do not change
\begin{document} % do not change

\begin{frontmatter} % do not change


\title{Estimation of shear viscosity based on transverse momentum correlations}


\author{Monika Sharma for the STAR Collaboration}

\address{Department of Physics \& Astronomy, Wayne State University, Detroit, MI, USA}

\begin{abstract}
Event anisotropy measurements at RHIC suggest the strongly interacting matter
created in heavy ion collisions flows with very little shear viscosity. Precise determination of
 ``shear viscosity-to-entropy'' ratio is currently a subject of extensive study \cite{Sean}.
 We present preliminary results of measurements of the evolution of transverse momentum correlation 
 function with collision centrality of $Au +Au$ interactions at $\sqrt{s_{NN}}$ = 200 GeV.
 We compare two differential correlation functions, namely {\it inclusive} \cite{Gary} and
 a differential version of the correlation measure $\tilde C$ introduced by Gavin ${\it et. al.}$ \cite{Sean,Pruneau}. 
 These observables can be used for the experimental study of the shear viscosity per unit entropy.
 \end{abstract}


\end{frontmatter}


%\begin{linenumbers}
\section{Introduction}
\label{intro}
Measurements of elliptic flow at RHIC (Relativistic Heavy Ion Collider) indicate based on comparisons with ideal hydrodynamics 
calculations that the quark gluon plasma produced in heavy ion collisions is a nearly perfect liquid 
\cite{sean1}. 
A measure of fluidity is provided by the ratio of shear viscosity to entropy density ($\eta/s$). 
Calculations based on Super-symmetric gauge theories \cite{Son} and uncertainty principle 
\cite{Gyulassy} suggest a lower bound, $\eta/s~\geq~1/4\pi$.
Elliptic flow has been the basic experimental probe for the estimation of $\eta/s$.  Based on recent 
measurements of elliptic flow and comparison with hydro models, the estimated range is $1<4\pi\eta/s<5$.
This suggests that the matter produced in $Au + Au$ collisions is indeed a low viscosity medium \cite{Teaney}.\\

In this paper, we present preliminary results of an alternative technique to determine the medium viscosity. The 
technique, proposed by Gavin {\it et. al.} \cite{Sean},
relies on measurements of the collision centrality evolution of transverse 
momentum two-particle correlation functions. This $\eta/s$ is estimated based on the longitudinal broadening of the correlations 
with increasing collision centrality. The broadening arises from longitudinal diffusion of momentum currents. It is quantitatively 
determined by the magnitude of the kinematic viscosity, $\nu = \frac{\eta}{Ts}$ (where ``T" stands for temperature), and the lifetime of the colliding system. We use 
differential extensions of the integral correlation observable $\tilde{C}$ proposed by Gavin {\it et. al} \cite{Sean}.\\

We present measurements of $\tilde{C}$ and inclusive ($\rho_{2}^{\Delta p_{1} \Delta p_{2}}$) as 
a function of the relative pseudorapidity and 
azimuthal angles of the measured particles. 
The observable $\tilde{C}$ is defined as
\begin{eqnarray}
\tilde C = \frac{{\left\langle {\sum\limits_{i = 1}^{n_\alpha  \left( {\eta _1 ,\varphi _1 } \right)} {\sum\limits_{i \ne j = 1}^{n_\alpha  \left( {\eta _2 ,\varphi _2 } \right)} {p_{\alpha ,i} \left( {\eta _1 ,\varphi _1 } \right)p_{\alpha ,j} \left( {\eta _2 ,\varphi _2 } \right)} } } \right\rangle }}{{\left\langle {n_\alpha  \left( {\eta _1 ,\varphi _1 } \right)n_\alpha  \left( {\eta _2 ,\varphi _2 } \right)} \right\rangle }} - \left( {\frac{{\left\langle {\sum\limits_{i = 1}^{n_\alpha  \left( {\eta _1 ,\varphi _1 } \right)} {p_{\alpha ,i} \left( {\eta _1 ,\varphi _1 } \right)} } \right\rangle }}{{\left\langle {n_\alpha  \left( {\eta _1 ,\varphi _1 } \right)} \right\rangle }}} \right)\left( {\frac{{\left\langle {\sum\limits_{j = 1}^{n_\alpha  \left( {\eta _2 ,\varphi _2 } \right)} {p_{\alpha ,j} \left( {\eta _2 ,\varphi _2 } \right)} } \right\rangle }}{{\left\langle {n_\alpha  \left( {\eta _2 ,\varphi _2 } \right)} \right\rangle }}} \right)
\end{eqnarray}
and {\it inclusive} is defined as
\begin{eqnarray}
\rho _2^{\Delta p_1 \Delta p_2 } \left( {\Delta \eta ,\Delta \varphi } \right) = \frac{{\left\langle {\sum\limits_{i = 1}^{n_\alpha  \left( {\eta _1 ,\varphi _1 } \right)} {\sum\limits_{j \ne i = 1}^{n_\alpha  \left( {\eta _2 ,\varphi _2 } \right)} {\left( {p_{\alpha ,i} \left( {\eta _1 ,\varphi _1 } \right) - \left\langle {p\left( {\eta _1 ,\varphi _1 } \right)} \right\rangle } \right)\left( {p_{\alpha ,j} \left( {\eta _2 ,\varphi _2 } \right) - \left\langle {p\left( {\eta _2 ,\varphi _2 } \right)} \right\rangle } \right)} } } \right\rangle }}{{\left\langle {n_\alpha  \left( {\eta _1 ,\varphi _1 } \right)n_\alpha  \left( {\eta _2 ,\varphi _2 } \right)} \right\rangle }}
\label{Eq13}
\end{eqnarray}

$n_{\alpha}(\eta_{i}, \phi_{i})$ represents the number of particles detected in an event $\alpha$ at pseudorapidity 
$\eta_{i}$ and azimuthal angle $\phi_{i}$. $p_{T, \alpha, i}(\eta_{i}, \phi_{i})$ stands for the transverse momentum 
of the $i^{th}$ particle in an event $\alpha$. $\left\langle {p_{T}(\eta _1 ,\varphi _1 )} \right\rangle$ is the 
average of the particle transverse momentum at $\eta_{i}$ and $\phi_{i}$ over the whole event ensemble.

\section{Analysis}\label{Ana}
This analysis is based on data recorded using the solenoidal tracker at RHIC (STAR) detector during the 
2004 data RHIC run at Brookhaven National Laboratory. $Au + Au$ collisions at $\sqrt{s_{NN}}$ = 200
 GeV were acquired with minimum bias triggers \cite{STAR}. 
This analysis is restricted to charged particle tracks from the STAR-Time Projection Chamber 
(TPC) in the momentum range 0.2 $< p_{T} <$ 2.0 GeV/c within the pseudorapidity acceptance of $|\eta|<$1.0. A nominal 
cut of distance of closest approach (DCA $<$ 3.0 cm) was applied in order to limit the selected tracks to primary 
charged particle tracks only. An event was accepted for analysis if its collision vertex lay within $|z|<$25 cm, where $z$
 stands for the maximum distance along the beam axis from the center of the TPC. The results reported here are based on 
 10 million minbias events.  We define centrality based on primary tracks within $|\eta|<$1.0. Centrality bins are calculated 
 as a fraction of the total multiplicity distribution.
 
\section{Results}\label{Results}
\begin{figure}[!htp]
\centering
\resizebox{5.25cm}{8.25cm}{\includegraphics{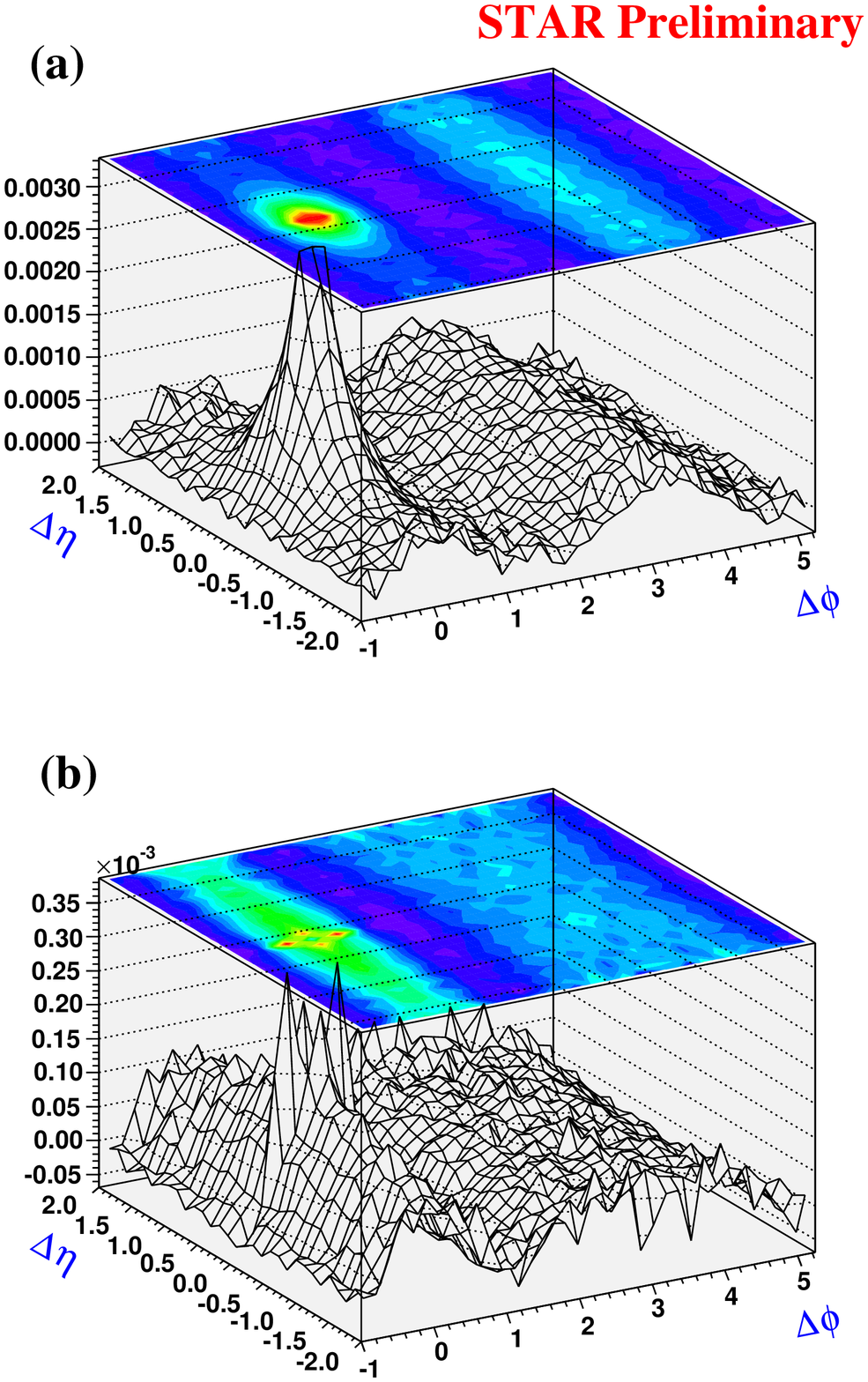}}
\resizebox{5.25cm}{8.25cm}{\includegraphics{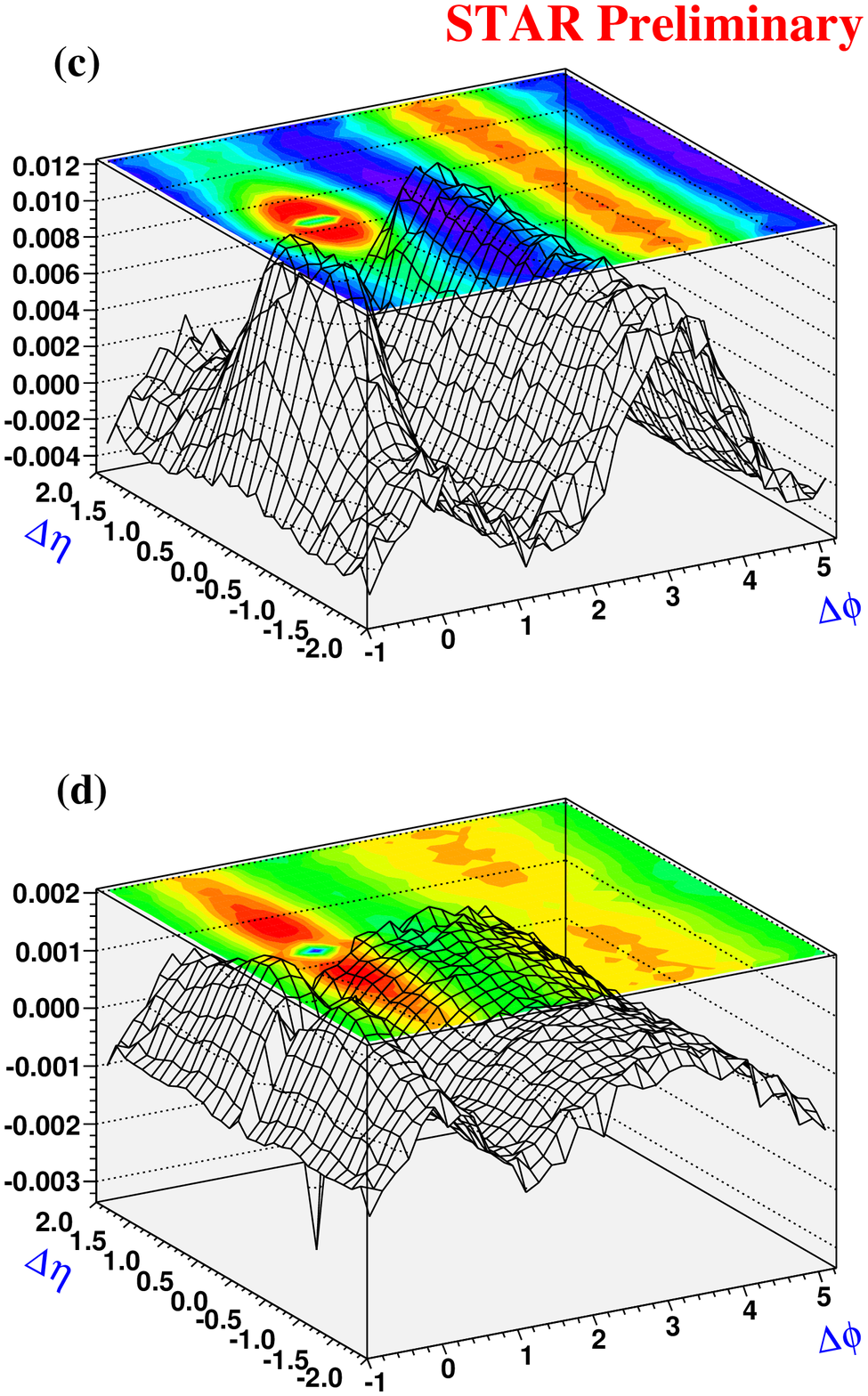}}
\caption[]{(Color online) (a, b) Correlation function, $inclusive$ ($\rho_{2}^{\Delta p_{T,1} \Delta p_{T,2}}(\Delta\eta, \Delta \phi)$) 
shown for 70-80 \& 0-5\% centrality, (c, d)  $\tilde{C}$ shown for 70-80 \& 0-5\% centrality in $Au +Au$ collisions at $\sqrt{s}$=200 GeV. 
These observables are plotted in units of $(GeV/c)^{2}$, and the relative azimuthal angle $\Delta \phi$ in radians.} 
\label{fig:1}
\end{figure}

Figures 1(a, c) and (b, d) show a comparison of $inclusive$ and $\tilde{C}$ correlations functions for 70-80\% \& 0-5\% centrality, respectively, 
 in $Au + Au$ collisions at $\sqrt{s_{NN}}$ = 200 GeV. The correlation function is 
plotted as a function of the particles' relative pseudorapidity, $\Delta \eta$, and azimuthal angles, $\Delta \phi$, 
using 31 and 36 bins, respectively.  The two  observables exhibit a ridge-like structure in the most central collisions (0-5\%) 
which is narrow in azimuth (near $\Delta \eta$ = 0) and extended over particles' relative pseudorapidity,
 $\Delta \eta$. In peripheral collisions both $\tilde{C}$ and $inclusive$ feature a near-side peak centered at $\Delta\varphi$=0, and a 
 broad away-side ($\Delta\phi \sim \pi$) ridge. The near side peak broadens progressively with centrality reaching a maximum
 in the most central collisions while the away-side amplitude progressively decreases from peripheral to central collisions.
 The $inclusive$ exhibits a single near-side peak structure whereas $\tilde{C}$ features a dip near $\Delta \phi \approx \Delta \eta 
 \approx 0$. The cause of this dip is under investigation.
We assume in this analysis that the broadening of the correlation function $\tilde{C}$ in $\Delta\eta$ is solely due to viscous diffusion 
effects and proceed to determine the evolution of the $\Delta\eta$ width with collision centrality. This is accomplished by fitting the $\Delta\eta$ 
projections of $\tilde{C}$ in the range $|\Delta\phi|<$1.0 radians.
Figures 2(a, c) and (b, d) show $\Delta\eta$ projections for $|\Delta\phi|<$1.0 radians for $inclusive$ and $\tilde{C}$ correlations functions 
for peripheral (70-80\%) and central (0-5\%) collisions, respectively. We parameterize the projections with a 5-component model. 
A wide Gaussian approximates the overall shape of the correlation function.
\begin{equation}
\tilde C\left( {b,a_w ,\sigma _w ,a_n ,\sigma _n } \right) = b + a_w \exp \left( { - \Delta \eta ^2 /2\sigma _w^2 } \right) + a_n \exp \left( { - \Delta \eta ^2 /2\sigma _n^2 } \right)
\label{Eq15}
\end{equation}
where $a_{n}$ and $a_{w}$ are the amplitude of the narrow and wide Gaussians, respectively. Similary $\sigma_{n}$ and $\sigma_{w}$ 
are the widths of the narrow and wide Gaussians. ``b" stands for baseline in Eq. \ref{Eq15}. 
 Widths obtained for peripheral ($\sigma_{w,70-80\%}$) and central ($\sigma_{w,0-5\%}$) collisions 
for $\tilde{C}$ are 0.53$\pm$0.01, 1.3$\pm$0.4, respectively.
Assuming the shear viscosity dominates the broadening of the correlation function for increasing system life times, the following
expression provides an estimate of the viscosity:
\begin{equation}
\sigma _c^2  - \sigma _p^2  = 4\upsilon \left( {\tau _{f,p}^{ - 1}  - \tau _{f,c}^{ - 1} } \right)
\label{Eq16}
\end{equation}
where $\tau _{f,p}^{ - 1}$ and $\tau _{f,c}^{ - 1}$ stand for the freeze-out time estimates in peripheral and central collisions. $\sigma_{c}$ and $\sigma_{p}$ represent the 
width of the correlation functions in the central and peripheral collisions.  
\begin{figure}
\centering
\resizebox{5.25cm}{8.25cm}{\includegraphics{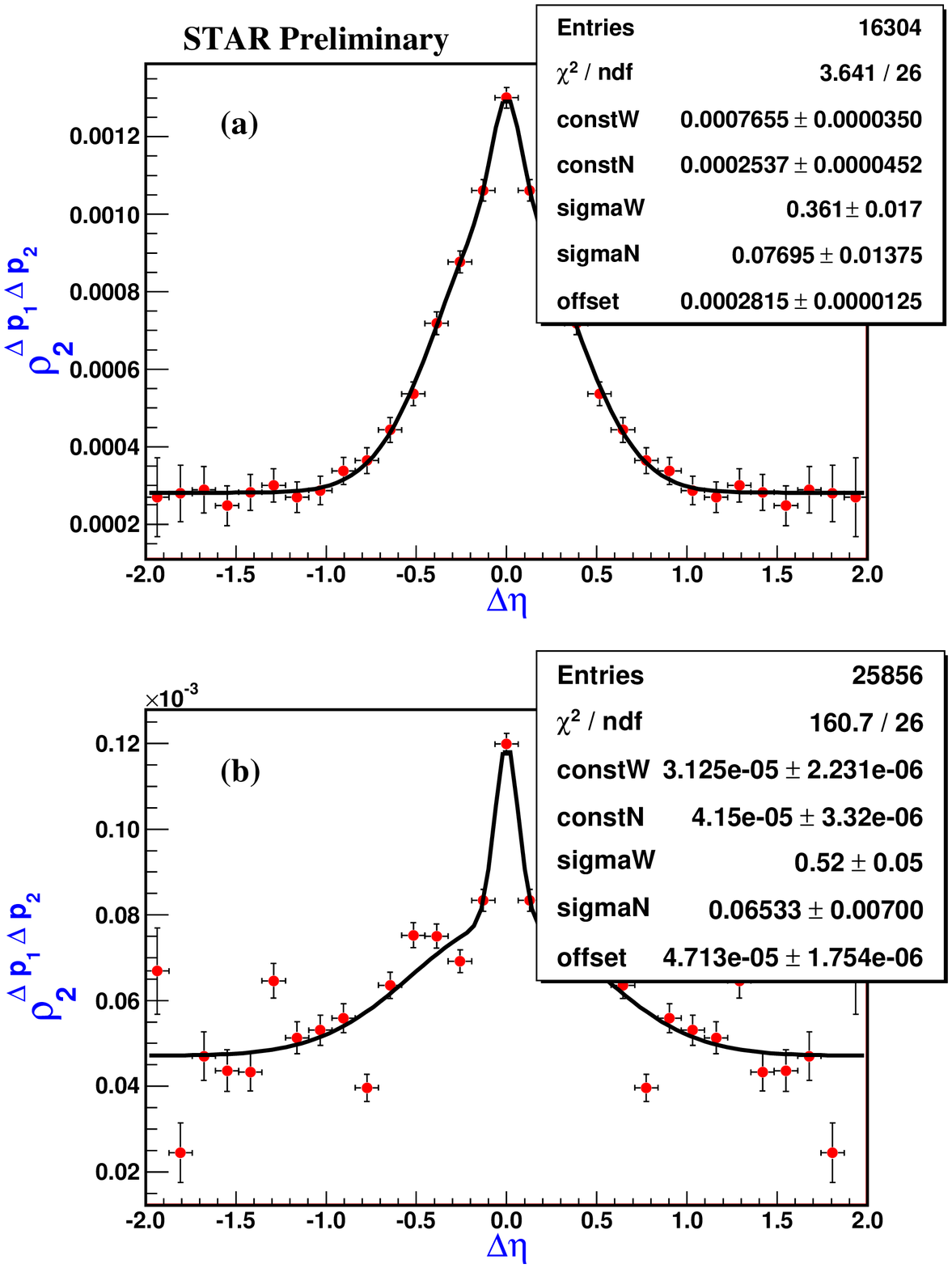}}
\resizebox{5.25cm}{8.25cm}{\includegraphics{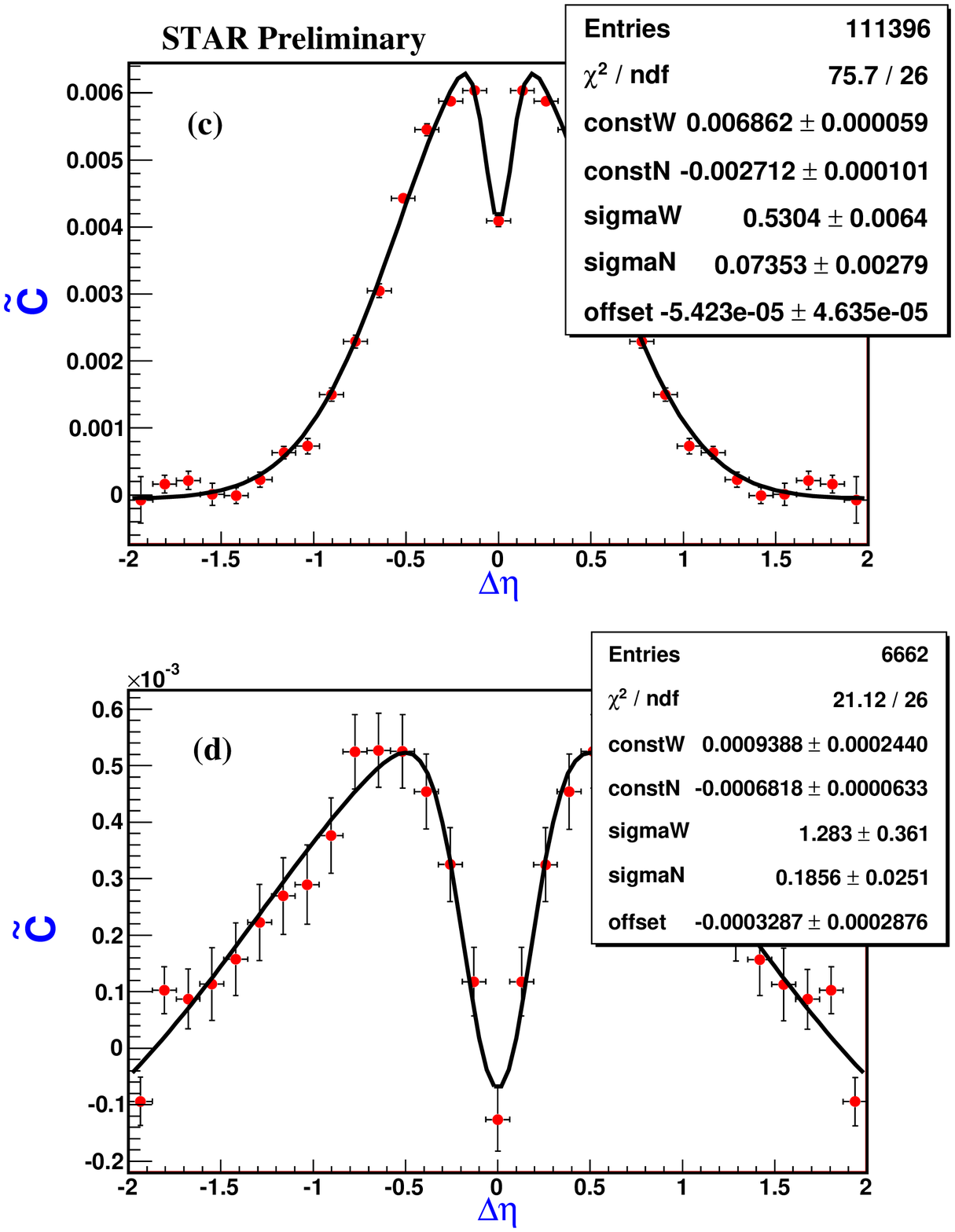}}
\caption[]{(Color online) Projection of $|\Delta\phi|<$1.0 radians on the $\Delta\eta$ axis (a, b) $inclusive$, $\rho_{2}^{\Delta p_{1} \Delta p_{2}}(\Delta\eta, \Delta \phi)$ correlations function for 70-80\% \& 0-5\% centralities, respectively, (c, d) $\tilde{C}$ shown for same centralities in $Au +Au$ collisions at $\sqrt{s}$=200 GeV. These observables are plotted in units of $(GeV/c)^{2}$.} 
\label{fig:1}
\end{figure}
\section{Summary}
We presented measurements of differential transverse momentum correlation functions in order to estimate the value of $\eta/s$
 based on the model by Gavin ${\it et. al.}$ \cite{Sean}. 
 The determination of $\eta/s$ will be sensitive to the freeze-out time estimate of the peripheral collisions posited by Gavin ${\it et. al.}$ \cite{Sean}.
 However, STAR measurements \cite{HBT} indicate a larger freeze-out time in peripheral collisions than that used in Ref. \cite{Sean}.

%\end{linenumbers}

\begin{thebibliography}{00}
\bibitem{Sean} S. Gavin and M. Abdel-Aziz, Phys. Rev. Lett. {\bf 97} (2006) 162302.
\bibitem{Gary} J. Adams {\it et al.} (STAR Collaboration), Phys. Rev. C {\bf 72} (2005) 044902.
\bibitem{Pruneau} M. Sharma and C. A. Pruneau, Phys. Rev. C {\bf79} (2009) 024905.
\bibitem{sean1} I. Arsene {\it et al.} (BRAHMS Collaboration), Nucl. Phys. A {\bf 757} (2005) 1;
K. Adcox {\it et al.} (PHENIX Collaboration), {\it el al.} {\it ibid.} 184; 
B. B Back {\it et al.} (PHOBOS Collaboration), {\it ibid.} 28; 
J. Adams {\it et al.} (STAR Collaboration), {\it ibid.} 102.  
\bibitem{Son} P. Kovtun, D. T. Son and A. O. Starinets, Phys. Rev. Lett. {\bf 87} (2001) 081601.
\bibitem{Gyulassy} P. Danielewicz and M. Gyulassy, Phys. Rev. D {\bf 31} (1985) 53.
\bibitem{Teaney} D. Teaney, Phys. Rev. C {\bf 68}, (2003) 034913; P. Romatschke and U. Romatschke, 
Phys. Rev. Lett. {\bf 99} (2007) 172301; H. Song and U. W. Heinz, Phys. Rev. C {\bf 77} (2008) 064901;
H. J. Drescher, A. Dumitru, C. Gombeaud and J. Y. Ollitrault, Phys. Rev. C {\bf 76} (2007) 024905.
\bibitem{STAR} M. Anderson {\it et. al.} Nucl. Instrum. Meth. {\bf A 499} (2003) 624; 
\bibitem{HBT} J. Adams {\it et. al.} (STAR Collaboration) Phys. Rev. C {\bf 71} (2005) 044906.
\end{thebibliography}
\end{document}